\documentclass[reprint,pre,letterpaper,superscriptaddress,amsmath,amssymb,aps] {revtex4-1}
\usepackage[utf8]{inputenc}
\usepackage{graphicx}
\usepackage{rotating}
\usepackage{amsmath}
\usepackage{threeparttable}
\usepackage{wrapfig}

\newcommand{\atBristol}{School of Biological Sciences, Univ. of Bristol, 24 Tyndall Avenue, Bristol BS8 1TQ UK.}
\newcommand{\atUH}{Dept. of Physics \& Astronomy, Univ. of Hawaii at Manoa, 2505 Correa Rd., Honolulu, HI 96822.}

\newcommand{\vvec}{\mathbf{v}}

\newcommand{\beq}{\begin{equation}}
\newcommand{\eeq}{\end{equation}}

\begin{document} 
\bibliographystyle{apsrev4-1} 
\title{Evidence for nanocoulomb charges on spider ballooning silk}

\author{E.~L.~Morley}
\email{erica.morley@bristol.ac.uk}
\affiliation{\atBristol}

\author{P.~W.~Gorham}
\email{gorham@hawaii.edu}
\affiliation{\atUH}

\begin{abstract}
We report on three launches of ballooning Erigone spiders observed in a 0.9 m$^3$ laboratory chamber, 
controlled under conditions where no significant air motion was possible. These launches were elicited by vertical, 
downward-oriented electric fields within the chamber, and the motions indicate clearly that negative electric charge 
on the ballooning silk, subject to the Coulomb force, produced the lift observed in each launch. We estimate the 
total charge required under plausible assumptions, and find that at least 1.15 nC is necessary in each case. 
The charge is likely to be non-uniformly distributed, favoring initial 
longitudinal mobility of electrons along the fresh silk during extrusion. These results demonstrate for the first time 
that spiders are able to utilize charge on their silk to attain electrostatic flight even in the absence of any aerodynamic lift.


\end{abstract}

\maketitle

\section{Introduction}

The phenomenon of aerial dispersal of spiders using strands of silk often called gossamer was 
identified and studied first with some precision
by Martin Lister in the late 17th century~\cite{MartinLister}, followed by Blackwall in 1827~\cite{Blackwall1827},
Charles Darwin~\cite{Darwin} on the {\it Beagle} Voyage, 
and a variety of investigators since~\cite{Duffey56, M3,M4,M5}. 
In modern parlance the behavior is known as
{\em ballooning}. This term evokes a central question:
does this mode of spider dispersal involve buoyancy forces, as the term suggests, or is 
it just a random process of aerial drift? Scientific investigation in the last several decades 
has largely dismissed the former possibility.
Here we present new evidence indicating that electrostatic buoyancy
is a real and potentially important component of spider ballooning dispersal.

\begin{wrapfigure}{r}{0.25\textwidth}
\vspace{-0.15in}
\centerline{\includegraphics[width=1.75in]{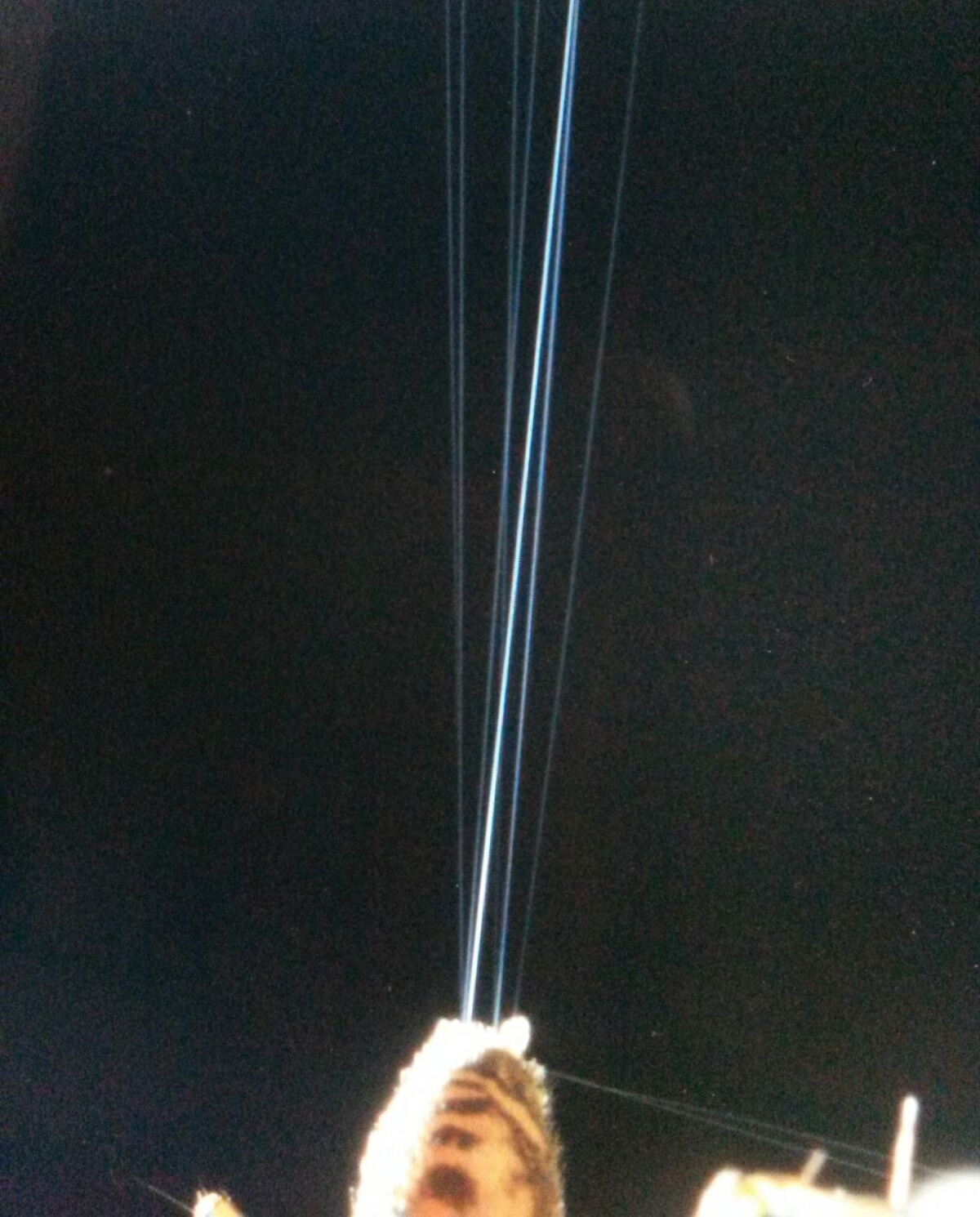}}
\caption{ \footnotesize \it Close-up image of a ballooning silk fan observed
during activities related to the experiment reported here. (Photo: M. Hutchinson.)
\label{silkfan} 
}
\vspace{-0.1in}
\end{wrapfigure}

Development of the physics basis of spider ballooning to date has focused on the hypothesis that 
ballooning was an exclusively aerodynamic process, relying on lift generated through complex
interaction between ballooning silk and the fluid dynamics of convective and wind-driven 
turbulence in the air.  Humphrey (1987)~\cite{Humphrey87} was the first to model ballooning as a fluid dynamic process, 
using a sphere (the spider)
suspended by a rigid rod (the silk), achieving some success estimating observed characteristics of ballooning. 
Further
refinements of the fluid dynamics approach have included flexible silk models~\cite{Reynolds2006,Reynolds2007}, and 
more sophisticated treatments of the effects of turbulence~\cite{Zhao2017}. These models do yield lift in
numerical simulations spider ballooning, but still appear to require significant upward components to the local wind velocity
distribution; whether actual wind momentum spectra provide the required distributions is still unproven,
particularly for takeoff conditions. Even so, recent detailed observations of spider
ballooning analyzed exclusively in terms of aerodynamic forces~\cite{Cho2018}
provide plausible evidence that larger spiders can use multi-thread fans of relatively long silk,
3~m or more, to achieve takeoff and buoyancy in low winds within a certain turbulence regime.
An example of multi-thread silk extrusion, shown in close proximity to the spinneret in
{\it Erigone spp.}, is shown in Fig.~\ref{silkfan}.

Curiously, all of the aerodynamic models fail in one regard: they 
provide no mechanism to avoid entanglement of silk during the takeoff and float period.
This issue becomes more acute for multi-thread silk fans, which are observed to splay out in
a triangular pattern, even when potentially hundreds of threads are involved~\cite{Schneider2001}.
Multi-thread fans have been noted throughout the history of spider ballooning
observations, and in all cases, observers are struck by the propensity of the silk fan to retain
its shape and order~\cite{Darwin,Schneider2001}, a fact that remains in tension with the putative turbulence required for buoyancy.

Under these conditions investigators have speculated that
electrostatic charge on silk plays a role in avoiding entanglement, and producing the observed
fan-like silk structures when multiple threads are emitted before launch~\cite{Schneider2001}. 
Polymer filaments extruded from capillaries in textile and other filament processing applications 
produce charges on the filament due to the phenomenon of flow electrification~\cite{Perolo2017},
thus some level of charge deposition on spider silk may be expected, and may
explain the silk's propensity to avoid entanglement.
The molecular complexity of the spidroin dope compared to simple non-polar hydrocarbons
may also increase the possible range of electrification in spider silk, but as our results
will show, if flow electrification is the source of the charge, it appears
to be under rather tight control of the spider during the spinning stage.

Electrostatic forces are actively produced by Uloborus spiders when spinning fluffy cribellate silk~\cite{Kronenberger2015}. 
This primitive type of silk forms puffs which have long been 
thought to hold their fluffy shape due to the repulsion of similarly charged threads, much like ballooning silk fans. 
There is thus evidence for spider control of the silk charge-state during spinning,
both in the development of silk fans among ballooners, 
and in Uluboris' capture silk, but to date there are no
direct measurements of the quantity of charge present in either of these cases.

The possibility of Coulomb force interactions in spider ballooning was considered
with regard to the Earth's global electric field, along with
considerations of the role of flow electrification in silk spinning, by Gorham~\cite{Gorham}.
However no experimental investigation of the potential
effect of the Earth's field in spider ballooning was undertaken until recently.  
In 2018, motivated by studies of the plausibility of physics underpinning such effects~\cite{Gorham},
and by the indications noted above that electrostatic fields may play some role in silk extrusion,
Morley and Robert~\cite{Morley2018} found the first experimental confirmation in a 
laboratory setting that electric fields elicit ballooning behavior in spiders.
These observations suggest that the Earth's electrostatic field~\cite{M7,Wilson1920,M6},
with a base value of typically 130~V/m, but with large variations in strength due to atmospheric
activity, may play a role in producing lift utilized by spiders during ballooning.
If so, spiders would be the first organisms known to make use of the Earth's field for dispersal,
or any other behavioral activity.

We report here further investigation of the role of charged silk in spider aerial dispersal, 
using measurements of three carefully observed ballooning
launches filmed within a laboratory chamber designed to be devoid of any air motion, containing
a vertical electric field determined by internal parallel conducting plates at the top and bottom of the chamber. These launches
were selected from a large range of ballooning-related behaviors that were elicited by the
presence of the vertical electric field, as shown in the controlled, blind study previously reported.~\cite{Morley2018}.
Analysis of these data demonstrates that the only possible source of the acceleration observed in these
events is the Coulomb force acting on charge contained on or entrained within the ballooning silk.

\section{Experimental methods}
%
\subsection{Morley \& Robert experiment.}
To provide context for the results reported here, we briefly review the previous
results reported by Morley \& Robert~\cite{Morley2018}.

\paragraph*{Test chamber.}
We use here data acquired in 2018 by Morley \& Robert with the same clear 0.9~m$^3$ plastic enclosure. 
The enclosure has a parallel plate configuration
to establish a vertical electric field, with careful thermal and humidity control preventing air currents. 
Two launch prominences were used,
each about 25 cm high with a 2~mm diameter tip,
one of non-conductive cardboard, and the other with conductive aluminum-foil.
Conductive 0.8 m square plates were positioned at the top and bottom of the chamber with 
0.8~m vertical separation. A 37~cm diameter plastic dish filled with water around the base of
the launch prominence prevented spiders from crawling away.

\subsubsection{Video \& experiment protocol.}
In this 2018 experiment,  
a group of 38 {\it Erigone} spiders, from a species known to be prolific ballooners, were tested 
for electric field response using the cardboard launch prominence.
The spiders were subject to a series of tests with 0V (control), 1000V, and 5000V plate potentials.
These plate potentials produce electric fields in the chamber, approximately 1250 V/m  and 6250 V/m, 
that are much larger than the Earth's fair-weather electric field, which typically ranges from 120-140 V/m. The higher values
were chosen for the test as representative of disturbed weather fields in the locale from which the spiders originated.

During the tests, spider behavior was logged using video, and then scored blind according to two behaviors closely
associated with ballooning: dragline drops from the prominence, followed by extrusion of ballooning silk, 
and tiptoeing, which is also followed by silk extrusion.
Each spider received
three trials, with voltage being switched on for a 2 minute interval during the trial, after a 5 minute initial
acclimatization period.
In each case, the launch site was carefully cleaned to remove any possible
cues that might transfer between tests. The voltage sequences were randomized to avoid any patterns,
and were unknown to the video observers. Each spider was also presented with only one treatment per day.

\subsubsection{Results.}
\begin{figure}[h!]
\includegraphics[width=3.4 in]{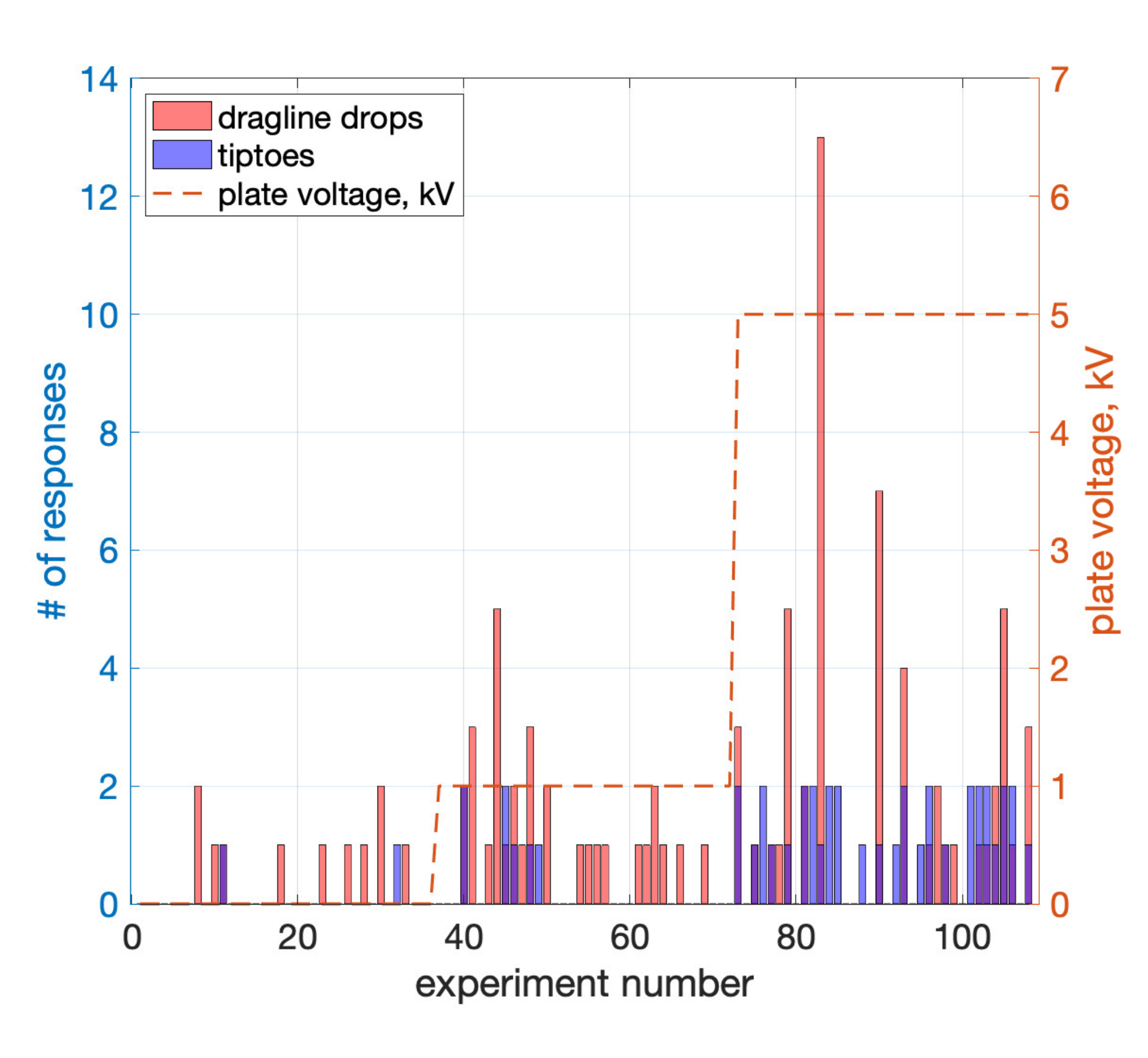}
\caption{ \footnotesize \it Frequency of two well-documented pre-ballooning behaviors
as a function of external plate voltage in the spider test chamber.
\label{Morleydata} 
}
\end{figure}

The results of this testing showed a strong propensity for ballooning behavior with 
increasing electric field strength~\cite{Morley2018}.
We summarize these results again in Fig.~\ref{Morleydata}, where the test sequence has been
ordered according to the voltage used (although in practice it was a random sequence). 
The correlation of ballooning-related behavior to the increasing
plate voltage is clear and statistically compelling, with a final $p$ value of $p<10^{-6}$,
close to $5\sigma$ in Gaussian statistics. 

This result from Morley \& Robert~\cite{Morley2018} provides the first experimental demonstration that
spiders initiate ballooning behavior in the presence of an electric field. This occurs in the
absence of any significant air motion, which is precluded by the closed, temperature- and humidity-controlled
chamber in which the tests were conducted.

Further tests using laser doppler vibrometry (LDV) on the same spider species also established the 
location of the spiders' sensory organs: the trichobothria on their metatarsal legs~\cite{Morley2018}. 
These elongated hairlike
features are known to be extremely sensitive to air currents~\cite{Barth2004}, 
the data also showed that the trichobothria response to E-fields is
distinguishable from that due to wind effects~\cite{Morley2018}.

\subsubsection{Spider behavioral data not previously analyzed.}
During these studies, a range of significant additional behavior relevant to ballooning was observed.
Categorizing and quantifying these additional activities were not part of the original controlled experiment.
To complement the work with the cardboard prominence, the aluminum-foil covered prominence was deployed,
to provide a more extreme case of the field concentration around a conductive prominence.
The increased electrostatic fields near the tip of the prominence led to an increase in ballooning-related activities,
which were also documented by video, but were not included in the original analysis. Analysis of these data
are presented as new results in this report.

The observed behaviors included 
\begin{enumerate}
 \item spiders extruding ballooning silk during high E-field periods, which became attached to the 
top or upper sides of the chamber, after which the spiders ascended the silk;
\item Dragline drops followed by
extrusion of silk, followed by partial lift by the silk of the spider in the field, while still attached by the dragline;
\item Actual ballooning launches after tiptoeing and silk extrusion.
\end{enumerate}
In all of these activities, it was qualitatively evident to observers 
at the time that electrostatic forces were clearly in play, and were associated with the ballooning silk,
not simply electrostatic charge accumulated on the spiders themselves.

Activities of type (1) occurred several times, and although it was evident that electric fields were likely
playing a role in extruding silk up toward the upper plate (as there was no other source of lift), there
was no clear way to quantify the amount of lift that was produced.

Activities of type (2) were also quite common, and in principle, if the spider size and mass, and the shape
of the catenary of the silk are known, it is possible to estimate the silk charge with reasonable precision.
However, the ballooning silk was fine enough that, while the observers could clearly distinguish it, the
video did not have adequate resolution to allow its shape to be quantified. 

In one case, a
spider that had extruded silk from a dragline drop then released the dragline and became free-floating in
the chamber, and remained so for several high-voltage on/off cycles~\cite{Morley2018}. The motion was captured by video,
and a portion of it is used to quantify the charge in that instance here.
Careful observation of this video also confirmed that the downward acceleration
during the voltage-off condition was not consistent with rappelling, indicating the silk remained free-floating
during this event.

\begin{figure}
\centerline{\includegraphics[width=3in]{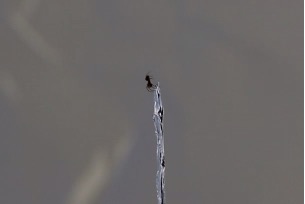}}
\vspace{2mm}
\centerline{\includegraphics[width=3in]{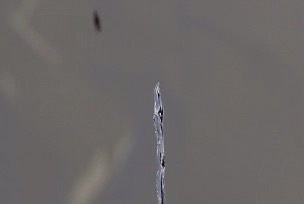}}
\caption{ \footnotesize \it Two examples of frames of the videos used for
the acceleration measurements. Top: tiptoeing spider just at the moment of launch,
with fine ballooning silk already deployed and extended vertically, as seen by observer although
unresolved in the video.
Bottom: Spider in motion, just before leaving the frame 160 ms later.
\label{frames} 
}
\end{figure}

Finally, two instances of activities of type (3), tiptoeing, followed by silk extrusion, followed by
a clearly observed launch, were videotaped. In both cases the spiders rose rapidly out of the field-of-view of
the video camera, which was then re-centered, following the spiders' motions. The spiders' vertical rise abated in both cases within a short
time due (evidently) to the impact of the silk with the upper plate, at which point the spiders
would continue climbing, or might rappel down. To ensure that we measured only acceleration
unaffected by either camera motion or the top plate, we used only those frames recorded before
the camera began to move to estimate the initial acceleration of the spiders, and from this, to derive the 
required charge, given the electric field in the chamber.

Two frames from one of the launches showing a spider in the tiptoe position just at the 
moment of launch and several frames later is shown in Fig.~\ref{frames}. The spider body length
is 2~mm. The motion was
not initially uniformly vertical as is evident in the frames, due to several-cm offsets in the silk axis relative
to vertical. For this analysis we use only the vertical component, since
motion in the horizontal direction is only measured in projection. 

The group of four spiders involved in this subset of behaviors were measured for mass and body size, but
unfortunately the data on individual spiders were lost, and only the value of 0.9 mg for the average mass of the four
spiders was preserved. Based on the video scales, we conclude that the variance of the group is relatively
small, but it does constitute an additional uncertainty, which we include by variational analysis below.

\section{Analysis framework}

In a uniform vertical electric field $\mathbf{E} = E_0 \hat{z}$, a spider of mass $m$, cross sectional 
area $A$, charge $q_b$ on the spider body, and total charge $q_s$ on the extruded silk, will experience the Coulomb force 
$\mathbf{F}_C = (q_b + q_s)\mathbf{E}$, and a gravitational force $\mathbf{F}_g = - m g \hat{z}$, with 
$g = 9.8~$m~s$^{-2}$.  Once the motion
begins, a drag force $\mathbf{F}_d$ will also develop.

The form of the drag force depends on whether the velocity of the spider is in the viscous drag regime, or 
in the pressure drag regime. In the viscous drag regime, we use Stokes drag as a model:
$$\mathbf{F}_{d,S} = 6\pi\eta r \mathbf{v}$$
and for pressure drag, 
$$\mathbf{F}_{d,P} = -\frac{1}{2} \rho C_d A v^2 \hat{v}$$
Here $\eta = 1.85 \times 10^{-5}$~kg~(m~s)$^{-1}$ is the dynamic viscosity of air at standard 
temperature and pressure (STP), $r$ is the equivalent spherical radius of the spider, 
$\rho = 1.2$~kg~m$^{-3}$ is the sea level density of air at STP, $\mathbf{v},v,\hat{v} $ 
are the velocity, speed, and unit velocity vector of the spider,
$C_d$ is the drag coefficient, and $A$ the cross-sectional area.

Which of these terms dominate will depend on the Reynolds number of the flow around the spider
$$R_e = \frac{\rho v d}{\eta}$$
where  $d$ is a characteristic dimension, $d \simeq \sqrt{A}$.  For the spiders considered here
$$R_e = 6.49 \left ( \frac{d}{1~{\rm mm}}\right  ) \left ( \frac{v}{10~{\rm cm/s}} \right )$$
and it is evident that except for the earliest part of the motion, $R_e > 1$,
pressure drag will prevail. Since Stokes drag may still play a role during the initial launch, we
do not neglect it in the model.
Spiders are not streamlined in their cross section,  so we expect
the drag coefficient to be $C_d \gtrsim 1$; in fact we will use experimental data
on spider free fall to estimate reasonable values of $C_d$ in the analysis below.

From Newton's second law, the sum of the forces is then
$$\sum_i \mathbf{F}_i =  m \mathbf{a} = \mathbf{F}_C + \mathbf{F}_g + \mathbf{F}_d $$
where $\mathbf{a}$ is the resulting acceleration. This equation is essentially the same
as Humphrey's 1987 result~\cite{Humphrey87} with the additional Coulomb and Stokes drag forces included, wind velocity $\vvec=0$,
and neglecting the $O(10^{-3})$ correction for the air displacement of the spider.
Inserting the forms for each of the terms and 
assuming all forces are acting only vertically as determined by the test chamber
$$m {a}  =   Q  E_0  - m g  - \frac{1}{2} \rho C_d A v^2 $$
where total charge $Q = q_b + q_s$.
We can solve this equation for $Q$ if the the acceleration is a measured quantity:
$$Q= E_0^{-1} \left [ m(a+g) + \frac{1}{2} \rho C_d A v^2 \right ] $$

The first term in brackets above, $m(a+g)$ is
dominated by the gravitational acceleration for any $a < 1$~m~s$^{-2}$. For spiders 
of mass  ${m} = 1$~mg
this term will have a magnitude $F \sim 10~\mu$N. For any spider with $d< 10$~mm, the drag force term will
be no more than 10\% of the gravitational term for any velocity below about 40~cm/s, well within the apparent 
range of velocities observed. So to first order, we can neglect the drag forces, and the resulting charge
equation becomes
$$|Q|= {[m(a+g)]}/{ |E_0|}$$.
Again since the observed accelerations $a << g$, the resulting charge depends to first order only on 
the spider mass, and for $|E| \simeq 7.5$~kV/m in the plateau region we find a lower limit on the 
total charge of
\begin{equation}
|Q| \geq \frac{mg}{ |E_0|}  = 1.3~{\rm nC} \left ( \frac{7.5~\rm{kV/m}}{E_0} \right ) \left ( \frac{m}{1~{\rm mg}} \right )
\label{Charge}
\end{equation}
The lower limit arises since the other terms initially neglected, drag, non-vertical motion, and the actual observed (non-zero) acceleration 
of the spiders, will lead to a larger charge.

The fraction of $Q$ that resides on the silk vs. the body of the spider is not yet determined. However,
measurements of induced charge on houseflies~\cite{houseflyTribo} and bees~\cite{beeCharge1,beeCharge2,beeCharge3} provides a scale for
the electrical capacitance of insect bodies, with total absolute charge from a wide variety of activities ranging
from 40-600~pC. Assuming spider carapace properties are electrically similar, the charge capacity
will be determined by the total area, For honey bees, this is estimated at 3.3~cm$^2$ on average~\cite{beeArea}. By comparison,
the spiders used in this experiment  are much smaller, with an estimated area of order $6 \times 10^{-2}$~cm$^2$.
We can expect the range of possible spider body charge to be $0.7~{\rm pC}~\leq q_b \leq 10~{\rm pC}$,
and thus the spider body charge will likely contribute less than 1\% of the total Coulomb force for spiders that attain lift.
We do not neglect the body charge in what follows, instead we will include both $q_s$ and $q_b$ into the final
fitted results below.

\begin{figure}

 \centerline{  \includegraphics[width=2.75in]{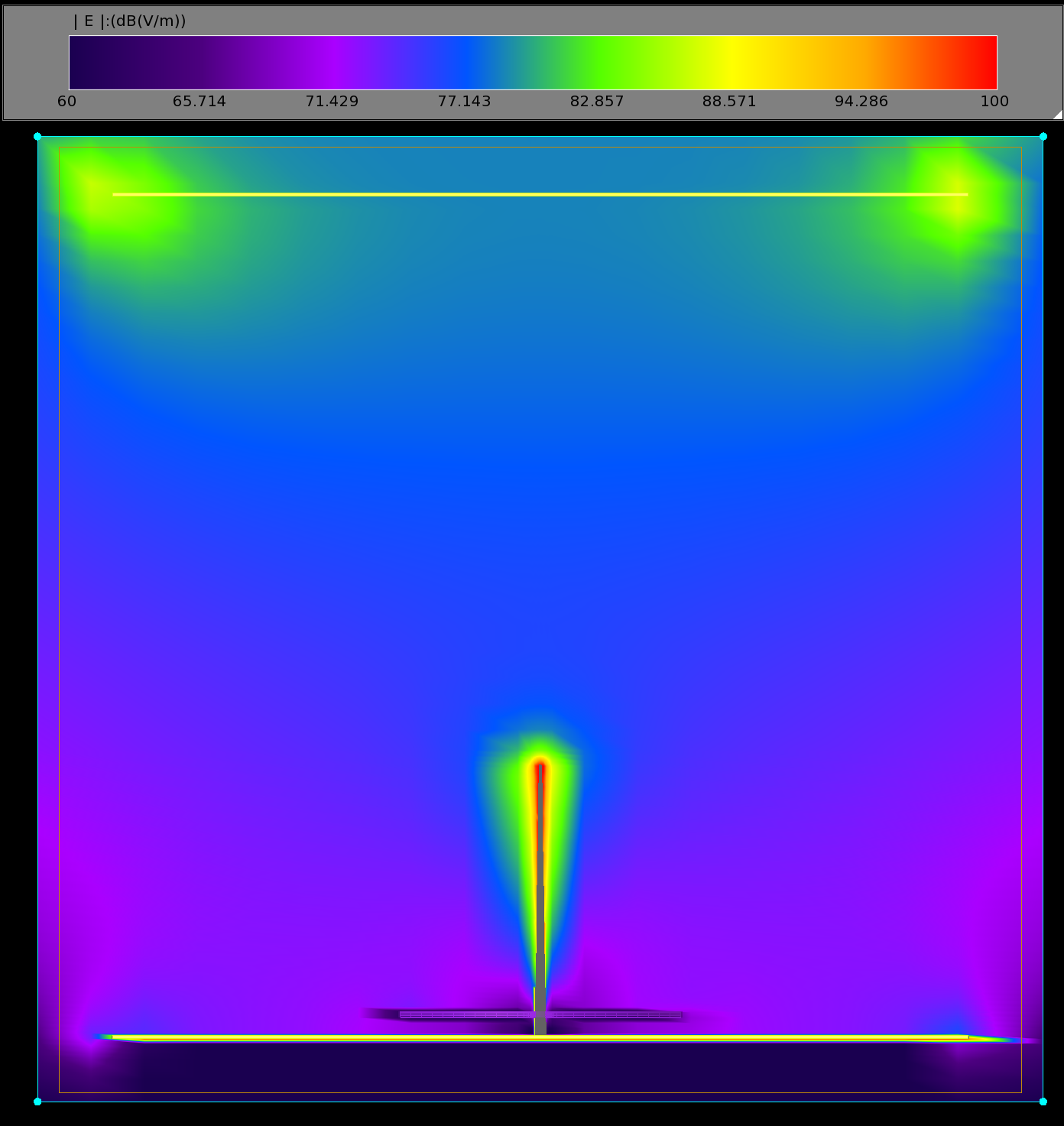}}
 \centerline{\includegraphics[width=3.5in]{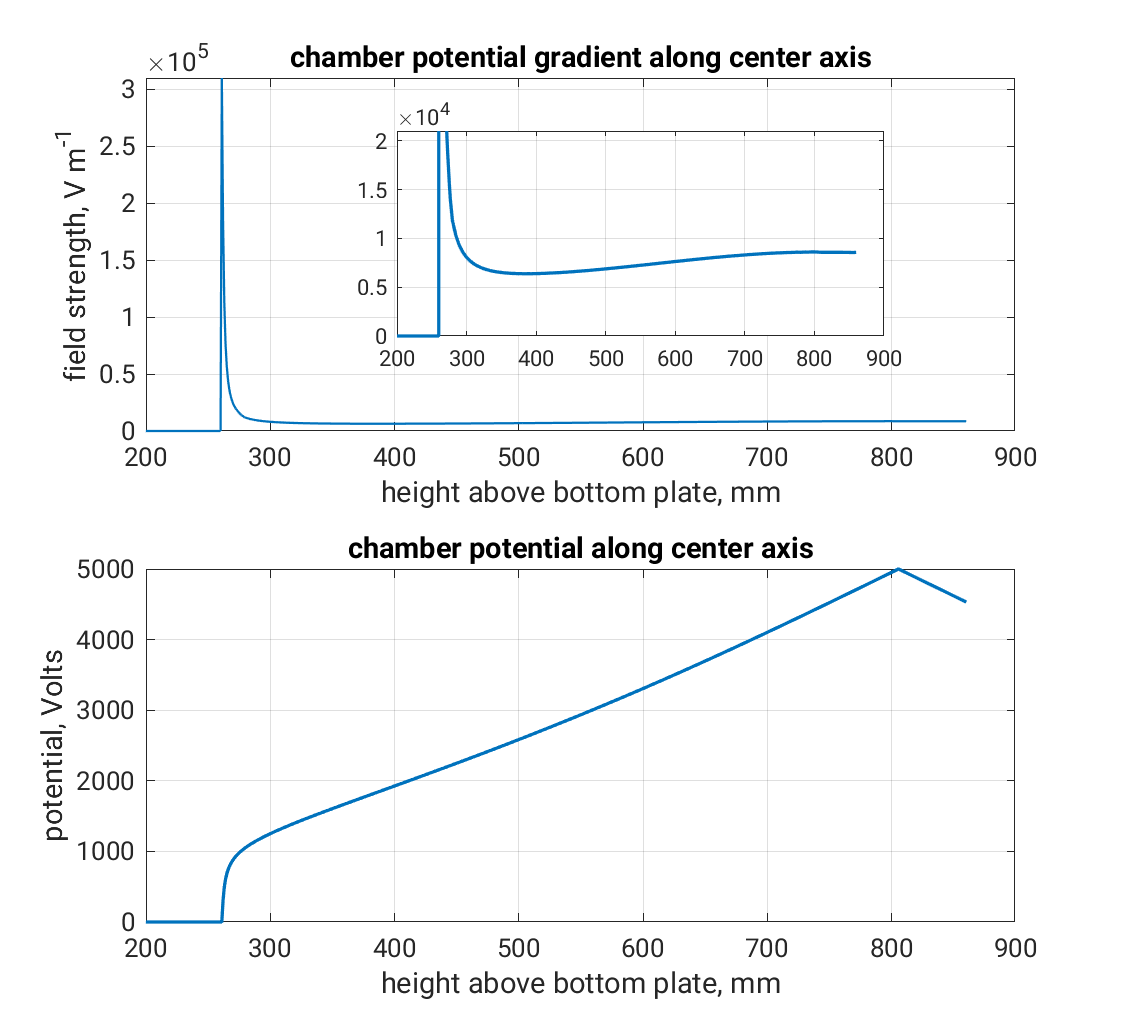} }
   \caption{\footnotesize \it
   Top: Electric field estimated by Remcom XFDTD model. Middle: vertical section through the
   modeled field above the tip of the conductive launch point. 
   Bottom: The integrated potential through the same section.}
   \label{Fields}
\end{figure}

Equation~\ref{Charge} sets the scale of the of the silk charge, but under idealized conditions. In practice, the
silk charge may be distributed and the field is non-uniform. To estimate the three-dimensional E-field configuration,
we used a commercial finite-difference-time-domain electrostatic solver, Remcom XFDTD, version 7.7~\cite{Remcom},
and the results are shown in Fig.~\ref{Fields}.
The system was modeled with precision in all significant details, except for the actual surface structure of the 
Aluminum foil prominence, which had a random texture due to the fabrication of the prominence by hand,
as evident in Fig.~\ref{frames}. Thus we expect that the model will accurately reproduce the fields on
mid- and large-scales within the chamber, but precise details in close proximity to the foil surface will
have increased uncertainty.

Under these E-field conditions, estimating the force requires an integral over
the charge distribution and the non-uniform field.  The net vertical force for a non-uniform field $E(z)$
and charge distribution $q(z)$ is given by
\begin{equation}
m {a}  =    \int_z^{z+L} q(z) E(z) dz  - m g  -  F_d
\end{equation}
where $F_d$ is the relevant drag force (which may be a combination of Stokes and pressure drag).
This equation can be written in the canonical form for coupled ordinary differential equations~\cite{CompPhys}:
\begin{align}
\frac{dv}{dt}  &=  \frac{1}{m} \int_{z(t)}^{z(t)+L} q(z) E(z) dz  - m g  - F_d \nonumber \\
\frac{dz}{dt}  &=   v(t) \label{ODE}
\end{align}
where the second equation
defines the coupling between vertical position and velocity. The charge distribution function $q(z)$ can be approximated
as a delta function at the spider location for the body charge, plus a distribution for the
silk charge, thus $q(z) = q_b \delta(z-z_b) + q_s(z)$.

Equations~\ref{ODE} are not straightforward to solve analytically,
and given that the charge distribution $q_s(z)$ is unknown, and the field distribution has no analytical model; 
it must solved numerically for assumed charge distributions. We consider two cases for $q_s(z)$,
(1) a uniform charge distribution per unit length of silk: $q_s(z) = (Q-q_b)/L$, where $Q$ is the total charge
and $L$ the length of the silk. This distribution may be expected for charge that is 
entrained or deposited along the silk, under conditions where the initial electrical conductivity 
of the silk is low enough that charge mobility can be neglected on the timescale of the launch.

At the other end of the scale is charge mobility that is high enough to allow charge to flow continuously
toward the distal end of the silk in response to the external electric field, given the intrinsic capacitance
and conductance of the silk as it is freshly extruded from the spinneret. For this case
(2) we assume a point charge $q_s(z) = (Q-q_b)\delta(z-z_s)$ located at the centroid of some finite 
segment of charge near the upper end of the silk, where the field is nearly constant.

We then 
evolve the equations of motion using a fourth order Runge Kutta (RK4) method to iteratively fit the position and
velocity for the three acceleration events we have observed. 

\section{Results}

\subsection{Video frame calibration.}

Video frame scales were calibrated by using the average spider
body length derived from the average 0.9 mg mass, estimated from statistical studies of spider
mass vs. body length for a large number of spiders of similar type~\cite{Penell2018}. The implied
mean body length for the spiders involved in the launches is $\sim2.0$~mm, 
consistent with the measured range of 1.8-2.8~mm of the full group of
38 spiders. We address the effects of this uncertainty later in this section.

\begin{figure*}[htb!]
\vspace{-0.1in}
\centerline{\includegraphics[width=6.5in]{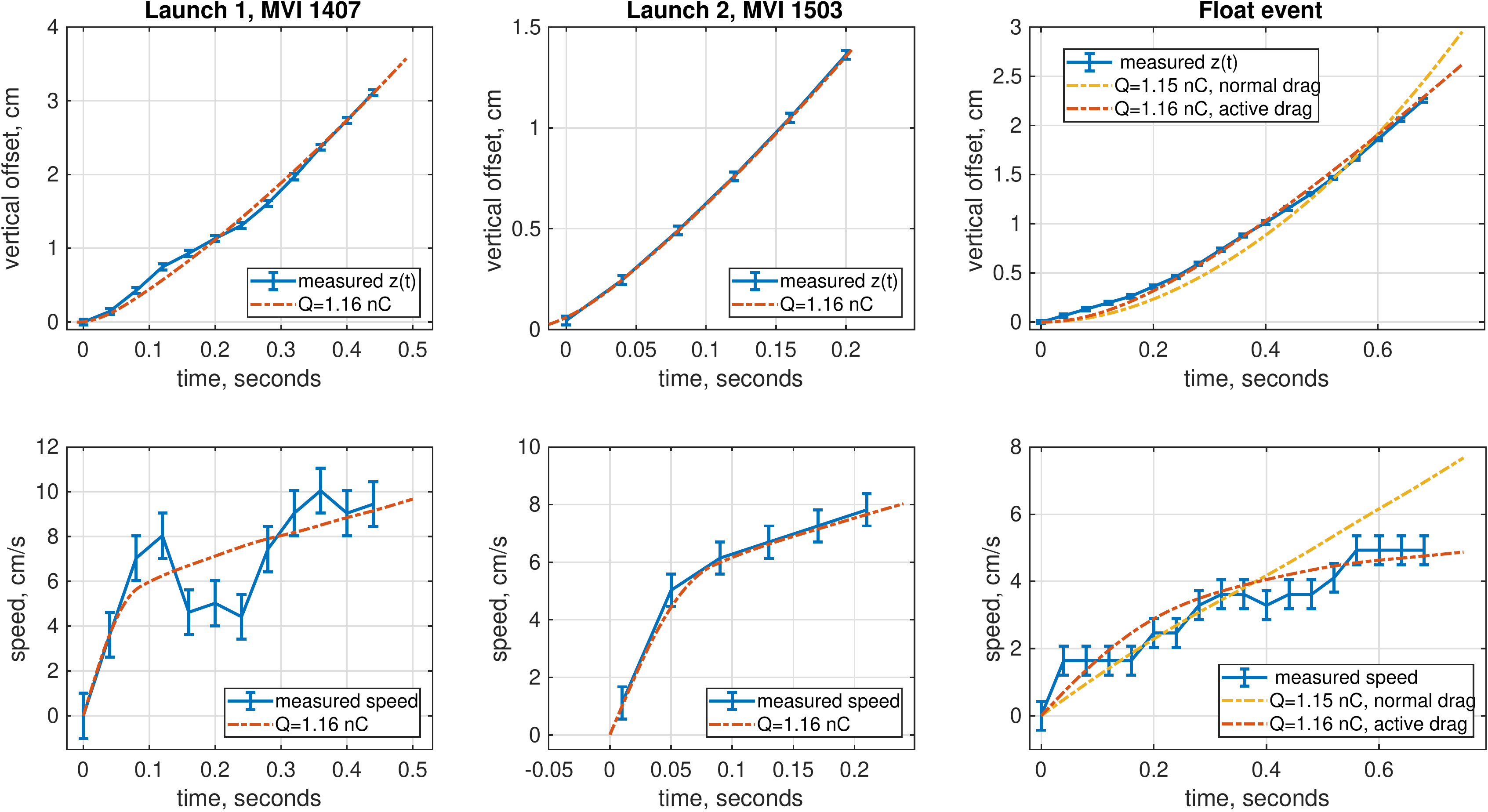}}
\caption{ \footnotesize \it Upper row: Measured and modeled positions vs. time for two
observed launches and the float event. Lower row, measured and modeled speeds.
\label{launches} 
}
\vspace{-0.1in}
\end{figure*}

Positions and their errors determined from centroids of the spider body position
for each frame. The frame rate is 25 frames per second, giving 40~ms per frame. 
For the force equation integrated using the RK4 method, we included gravity,
the Coulomb force, and both viscous and pressure drag. An additional term
accounting for stretch in the ballooning silk under tension just prior to launch was estimated
using a Hooke's law spring force, $F = k\Delta z$, with spring constant $k$ dropping to zero once
the initial strain was relaxed. The total range of motion that could be observed without losing
calibration due to camera motion was 1.5-3.5~cm. While a larger range of motion would certainly
be preferred, the standard errors on positions measurements were fractions of a mm, so 
this limited range of motion still provides clear constraints on the forces involved.

\subsection{Exclusion of uniform charge density.}

In our results, continuous uniform charge distributions produced solutions in which the spider oscillated
around the launch point with a period of about 0.5 seconds. This is due to the fact that the strong
local fields at the launch point, coupled with the charged silk in very close proximity, 
initially dominate the acceleration, forcing the fitted linear charge density to a relatively low value in order not to exceed the observed
speeds. Then, when the spider leaves
the immediate vicinity of the launch point, within less than 1~cm, the precipitous drop in the 
field causes the lift force due to the underestimated charge density to
fall below that of gravity. The spider then falls back down until the strong field region is entered.
This behavior is completely inconsistent with the observed acceleration, and we thus exclude a uniform distribution
of charge as a viable solution.

In fact, as noted above, we believe that
fresh ballooning silk allows for charge mobility high enough to ensure that the charge
migrates rapidly away from the strong local E-field at the launch point.
A recent study of the electrical conductivity of spider silk found it to be a very strong function of local relative humidity,
increasing by more than three orders of magnitude, from $<10^{-6}$ to $\sim 10^{-3}$ S/m over the range of 30\% to 70\% 
relative humidity, a remarkable change~\cite{silkCond1}.
These measurements were made with dragline silk tested well after harvesting; to our knowledge no measurement
of the electrical conductivity of fresh ballooning silk has been published.

We hypothesize that freshly extruded silk, which has only just been dehydrated within 
moments of leaving the spinneret, may have a relatively high conductivity during its 
initial phase. Under the high launch point field concentration, charge would naturally be
driven away from the launch point and toward the end of the silk, which would be
subject to far more uniform fields in the upper region of the chamber.
Our results are in fact consistent with either a concentrated charged region near
the silk upper end, or a distributed silk charge within the plateau region of the field,
as seen in Fig.~\ref{Fields}.

\subsection{Model fit results.}

The results of these RK4 model fits, along with the measured data, are shown in Fig.~\ref{launches}, for both
vertical position relative to the start point (upper frames), and the vertical speed (lower frames).
The non-linear nature of the motion is evident in each case, both in position, and speed.
The fitted
charge, with negligible statistical errors, is $q_s = 1.15-1.16$~nC, consistent with
equation~\ref{Charge} above for the average 0.9~mg mass of the spiders involved. 
The fitted value for the spider body charge is also consistent for all three data sets:
$q_b = 3\pm 0.5$~pC, which is of the same order of magnitude as the linear
charge density, about 10~pC/mm, implied by the electric field models at the upper end of
the aluminum foil prominence.

These were values determined
using a grid-search minimization of the $\chi^2$ function for each of the three
ballooning events:
\begin{align}
 \chi^2 ~~=~~ &\sum_{i=1}^N \frac{ (z_o(t_i) - z_m(t_i))^2 }{\sigma_z^2}~~~ \label{chi2} 
\end{align}
where $N$ is the number of frames, $z_o(t_i)$ is the observed vertical position at time $t_i$, 
$z_m(t_i)$ is the modeled position at time $t_i$, and $\sigma_z,~\sigma_u$ are the estimated standard errors in
position and speed, respectively. The model positions are given by
the numerical solutions to equation~\ref{ODE} above.

Statistical uncertainties are typically $\sigma_z = 0.4$ to  0.5~mm in the video frame position estimates, 
which translates to about $\sigma_u = 1$~cm/sec uncertainty in the speeds.
Since the speeds are derived from the positions, they do not provide
independent information for the $\chi^2$ minimization, but are shown in Fig.~\ref{launches}
because they show more clearly the transient effects: the local repulsion of the 
spider from the launch point due to the body charge, and the short duration of
the snap-back of the stretched thread, both of which contribute to the early higher
acceleration of the spider at launch. These effects are absent in the float event,
since the spider was already at least 6~cm away from the launch prominence when the acceleration started,
well outside of the highly enhanced field at the tip of the prominence.

In these events the spiders typically rose an additional distance of 20~cm or more beyond
the vertical range we analyze here, but the requirements for calibration of the 
scale of the motion lead us to restrict the data to the initial period before camera motion began. Despite 
this restriction, we emphasize here that this motion is completely inconsistent with
any other force available to the spiders. There was no thermal gradient to produce
any significant air motion within the chamber, certainly not at the speeds of 4-8~cm/s as
observed. Spiders were watched carefully to ensure that no silk had attached to
the chamber walls or top, and the motion observed in these launches involved no
climbing actions by the spider. The only known source of lift in this case is the Coulomb
force, and the observed motion requires that this force must be primarily exerted
on the upper portion of the silk, where we conclude the charge resides.

\subsection{Systematic uncertainties.}

Systematic uncertainties in these results include (a) the unknown drag coefficients of the spiders;
(b) possible pendulum motions immediately after launch due to off-axis 
E-field effects near the launch point; (c) uncertainties in silk strand length;
and (d) the uncertainty in the video size scale.

For (a) we used values consistent with spider free fall measurements
made by Suter (1992)~\cite{Suter92}, which imply surprisingly large drag
coefficients, $C_d \gtrsim 3$, based on estimated cross-sectional body areas,
perhaps because of the complex shape, including the legs, for which the effective area
is difficult to estimate. We varied the pressure drag coefficient $C_d$ 
by factors of two around our adopted baseline value $C_d = 3$ with no effect on the results.
Other systematics were also checked by variational methods, and the
resulting charge estimates were found to be quite robust to these variations.

As noted above, pendulum-like motions were observed in the early part of one of the launches;
we compensate for (b) by using only the vertical component, which is conservative
in that it underestimates the total acceleration. To address (c), an approximate
estimate of strand length $L \simeq 0.4$ to 0.5 m was obtained by observing at what
elevation the spider motion was abated, but we also did not
assume any strand-length dependent parameters, solving for total charge
under the assumption only that there was charge migration to its upper half, in the region where
the E-field plateaued, during 
extrusion. As we have also noted above, the observed motion is not consistent with a uniform charge
distribution. 

The effect of the scale uncertainty ((d) above), which we expected to translate linearly to
an uncertainty in the resulting charge, was in fact quite small, as we found by
fitting for the charge while varying the scale within the range of the uncertainty.
We attribute this to the fact that the upward spider vertical accelerations we observed, while clearly
evident, were still small in magnitude compared to the gravitational acceleration,
and thus errors in these only affect the resulting charge at second order, since to first order
(eg., equation~\ref{Charge}) the charge is independent of spider acceleration
for $a << g$.

\subsection{Quality of the model fits.}

For the two launches, the RK4 numerical model fits the overall motion well. The speed in the 1407 launch includes
some variation, probably due to some pendulum motion of the spider, that is not yet 
very well-modeled, due to lack of precise knowledge of the silk shape, but once the pendulum
motion abates, the model again aligns with the data.

For the float event, the observed
motion appears to include a deceleration in excess of the expected spider drag, and indicates
some deficiency in the model, although the fitted charge remains quite similar.
We thus also show the results of
an Ansatz model in which the spider pays out silk during the upward motion.
We model this as an additional drag-like term, which we denote as ``active drag.''  
In practice its magnitude is equivalent
to an increase of an order of magnitude in the pressure drag coefficient , but results in a negligible change
in the magnitude of the silk charge required by the fit.
This is illustrative but not conclusive -- it
remains one of the open questions of the resulting motion. Despite these moderate discrepancies the
measured motion shows clear and compelling evidence that nanocoulomb charges 
must reside on ballooning silk.

\begin{table}
\begin{threeparttable}[htb!]
\caption{Results of charge fitting for the three ballooning events reported here.}
\label{resultstable}
\centering
 \begin{tabular}{|c|c|c|c|c|}
 \hline
Ballooning  &  estimated & total   &$\chi^2_{min}$ &  DOF\tnote{c}~~\\ 
 event\tnote{a}   & charge, nC & error, nC\tnote{b}~~    &  &    \\ \hline
MVI1407 &  $1.155$ &  0.02 & 49.9 & 11\\
MVI1503 &  $1.158$ & 0.02  & 0.95 &  5 \\
Float, normal drag\tnote{d}~~ & $1.154$ & 0.02 & 544 & 17 \\
Float, active drag\tnote{e}~~ & $1.160$ & 0.02  & 76.7  & 17 \\ \hline
 \end{tabular}
 \begin{tablenotes}
  \footnotesize
  \item[a] Direct launches are indicated by the video recording number; the ``Float'' case involved a free-floating spider.
  \item[b] Combination of estimated statistical and systematic errors.
  \item[c] Degrees of freedom for the position $\chi^2$ as given in equation~\ref{chi2}.
  \item[d] Best estimate for simple Stokes + Pressure drag.
  \item[e] Assumes spider actively extrudes silk after launching to reduce acceleration.
 
 \end{tablenotes}
\end{threeparttable}
\end{table}

In Table~\ref{resultstable} we show a summary of the fitted charges for the three
ballooning events, with two drag scenarios considered for the final float event.
The table includes estimates for the combined systematic  + statistical
errors, along with the final $\chi^2$ and the number of degrees of freedom
in each case. In the float event the discrepancies in the fits are evident in the final
$\chi^2$ values; although the active drag is a much better fit to the data, the model
is still not fully commensurate with the data.
For MVI1503, the fit is reasonably good, and for MVI1407, the tension in the fit
comes mainly from fluctuations in the trajectory, likely due to pendulum motion.

These results represent the first direct measurement of such charges via acceleration
in a controlled environment. It is surprising that the values measured all fall within a very small
range. This raises important questions about the silk extrusion: do ballooners have the ability to
tune or even modify in real time amount of charge on their ballooning silk, depending on the
environment? If so, how is the charge manipulated? Is it embedded onto the silk by the
extrusion process itself, or entrained from charge at the launch point?

\begin{figure}
\vspace{-0.1in}
\centerline{\includegraphics[width=1.65in]{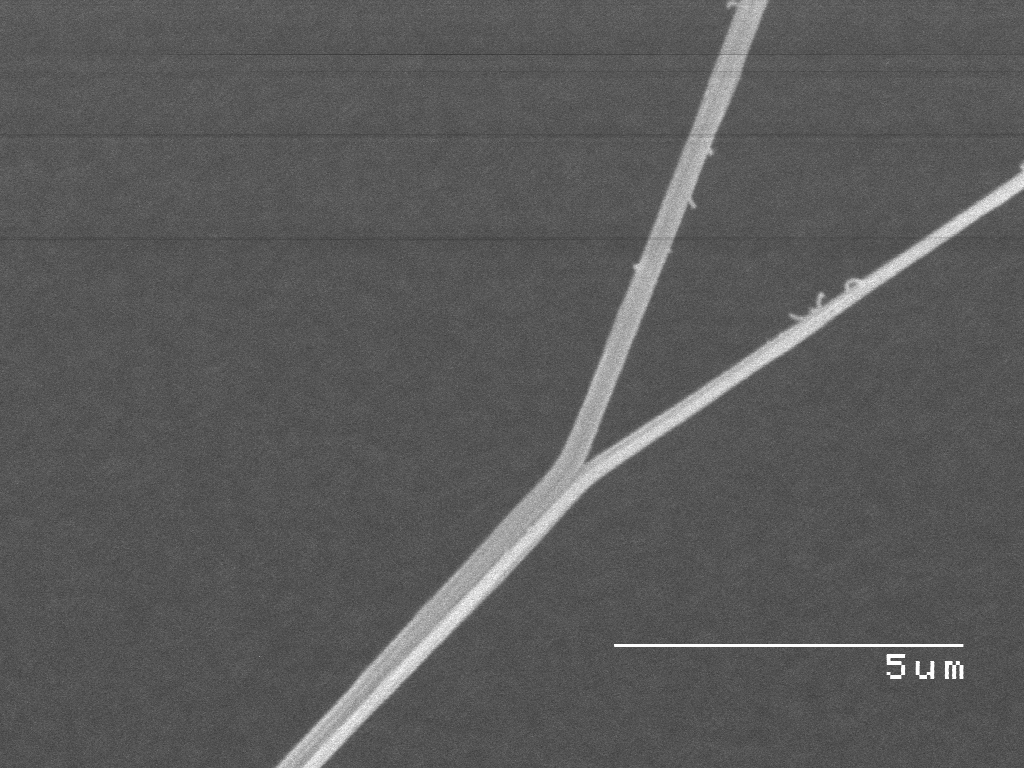}~~\includegraphics[width=1.65in]{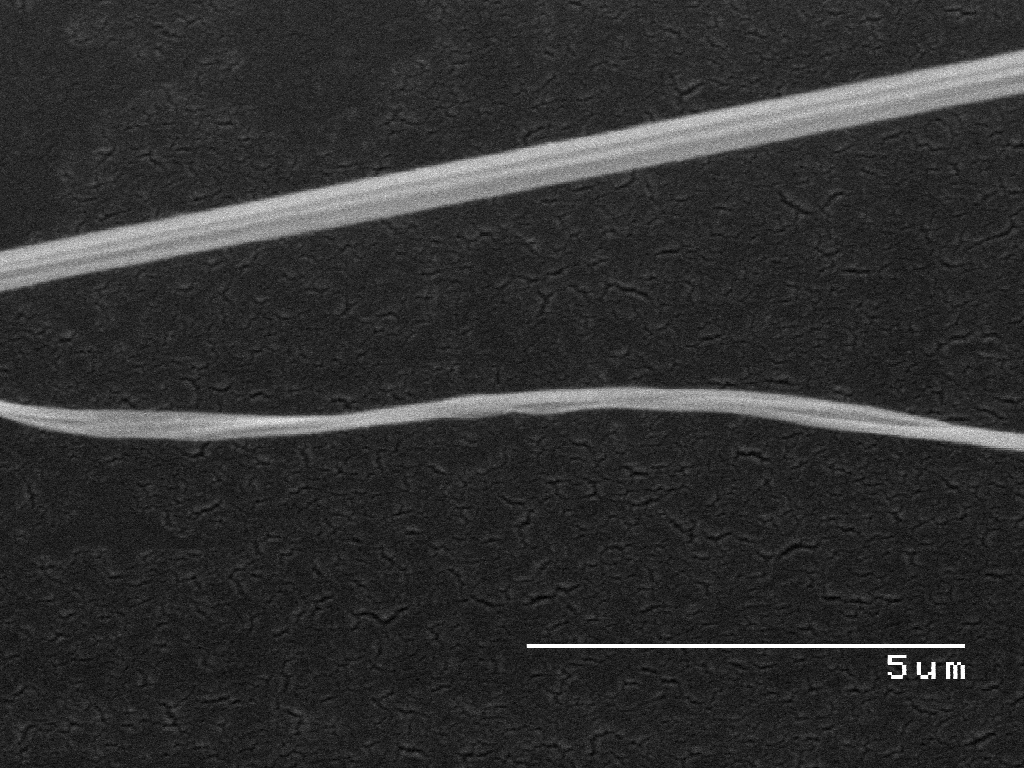}}
\caption{ \footnotesize \it Two SEM images from ballooning silk gathered during
the experiment activities reported here (credit E. Morley).
\label{silkSEM} 
}
\vspace{-0.1in}
\end{figure}

\subsection{Space charge issues.} 

The quantity of charge observed corresponds to about $7 \times 10^9$ excess electrons.
This implies that spider silk is a form of {\it electret},
a material able to store and retain free charge for some period long compared to the intial charging time. 
The observed diameter of the multi-stranded ballooning silk in this experiment was typically $\sim340$~nm, as confirmed
by scanning electron microscopy, examples of which are shown in Fig.~\ref{silkSEM}.
If we assume the excess electrons are concentrated uniformly in the 
upper 40\% of the silk length, in a segment with $\ell=20$~cm, the resulting space charge density 
at the nanoscale is $\rho = 1.1 \times 10^{-3}$~electron~nm$^{-3}$. We can then estimate the excess charge per spidroin
molecule. Following \mbox{Erickson~(2009)~\cite{Erickson09}}, the folded protein volume $V$ can be approximated as
\begin{equation}
 V ({\rm nm^3} ) \simeq 1.212 \times 10^{-3} ~{(\rm nm^3~Da^{-1} )} ~ \times ~M~{(\rm Da)}
\end{equation}
for molecular mass $M$ in Daltons. For a mean molecular mass of $\sim$300~kDa for spidroin, 
the resulting volume is 364~nm$^3$, implying 0.42 excess electrons per folded spidroin molecule.
This is a quite low charge excess for these large molecules, presenting no obvious problems given 
the typical $\sim$3500 amino acids constituents of silk spidroin.

However, the implied charge density of the silk, viewed as an electret in this case, assuming an effective radius
for the ballooning silk of $r_e \simeq 170$~nm, 
is $\rho = 64$~kC~m$^{-3}$. This is much larger than the bulk charge density of typical polymer
electrets which can reach 2.5~kC~m$^{-3}$~\cite{SpaceCharge98}. In fact, if the silk is
conductive enough to allow charge migration during the initial spinning of the ballooning thread,
a period of order a few seconds at most, it may be assumed that the charge would migrate to
the surface, and thus yield near-surface charge densities in even stronger tension with those seen in polymer
electrets.

There are several ways in which this tension may be mitigated. 
First, recent work has demonstrated
space charge densities as high as 100~kC~m$^{-3}$ in micro-electrets in the form of 500~nm diameter silica-based spherules~\cite{highSC2018}.
As the authors of this study demonstrate, such charge densities are much easier to accommodate in
micrometric scale bodies with low dimensionality, a feature that spider silk shares with this example.

Second, the authors of this study on micro-electrets also note that the structure of the spherules used, 
which involves a process that builds up the spherule from smaller silica nanoparticles, yields a higher internal
surface area for charge accumulation than a homogeneous material. The structure of spider silk, spun from
series of $\sim$100~nm-scale fibrils (as seen in Fig.~\ref{silkSEM}), also yields a significantly larger
surface area than the equivalent single-strand cylinder. The corresponding charge-trapping efficiency
of silk, with its complex amino-acid structure, is thus likely to be much higher than a normal bulk polymer electret.

Finally, while the conductivity required of the fresh silk in order for space charge to propagate longitudinally
to the distal end of the strand (as required by our observations) is much higher than for normal dielectrics, 
it is still far below that of metals. The conductivity
range for silk as noted above, from $<10^{-6}$ to $\sim 10^{-3}$ S/m over the range of 30\% to 70\% 
RH~\cite{silkCond1}, falls in the transition region appropriate for semiconductors; thus for example, 
the conductivity of amorphous silicon can be as low as $10^{-3}$ S/m, and that of cadmium sulfide as low
as $10^{-5}$ S/m. Space charge effects in such materials are common and in fact help to define the
nature and behavior of such materials~\cite{spaceCharge}. 

It is worth also noting that the spinning process itself, in which the final stage of motion of 
the spidroin dope through the spinneret involves dehydration of the dope via ionic effects at the
walls of the spinneret tubule, may yield a material with a much higher conductivity close to the 
central axis of the fibroin than at its perimeter. As noted in the introduction, 
flow electrification is one mechanism for creating the charge, but in the spinning of simple
polymeric fibers, there is no equivalent dehydration phase. Flow electrification with
a uniform conductivity typically
produces a radial profile with space charge concentrated near the walls of the flow tube~\cite{flowElect, flowElect2};
in the case of spinneret flow, a conductivity gradient could significantly modify the result.

Such an asymmetric conductivity profile could
in fact yield enhanced longitudinal migration of the space charge, while limiting surface charge density
due to the decreased mobility near the perimeter. Under these conditions, where the hydration
state of the silk is closely coupled to its conductivity, the relative permittivity $\epsilon_r$ is also likely to
have a strong radial dependence, since $\epsilon_r \sim 80$ for water. Such a permittivity
gradient would also tend to reduce the radial space charge mobility due to the polarization of
the silk, which would reduce the apparent radial electric field.

In summary then, while the space charge density implied by our observations are quite high,
further studies of the nanoscale structure, permittivity, and conductivity of freshly spun
silk are necessary to understand the physical nature, nanoscale and macroscale distribution, and mobility of the charge.

\section{Implications for fair weather field conditions.}

The electric field conditions used in these experiments was based on conditions observed
in active or disturbed weather, with ambient fields that are much stronger than the 
Earth's fair APG of $\sim 130$~V~m$^{-1}$. One of the important questions this raises is
whether spiders make use of the fair weather field at all, or do they only make use of
silk charge under high APG conditions.

\begin{figure}
\centerline{\includegraphics[width=2.75in]{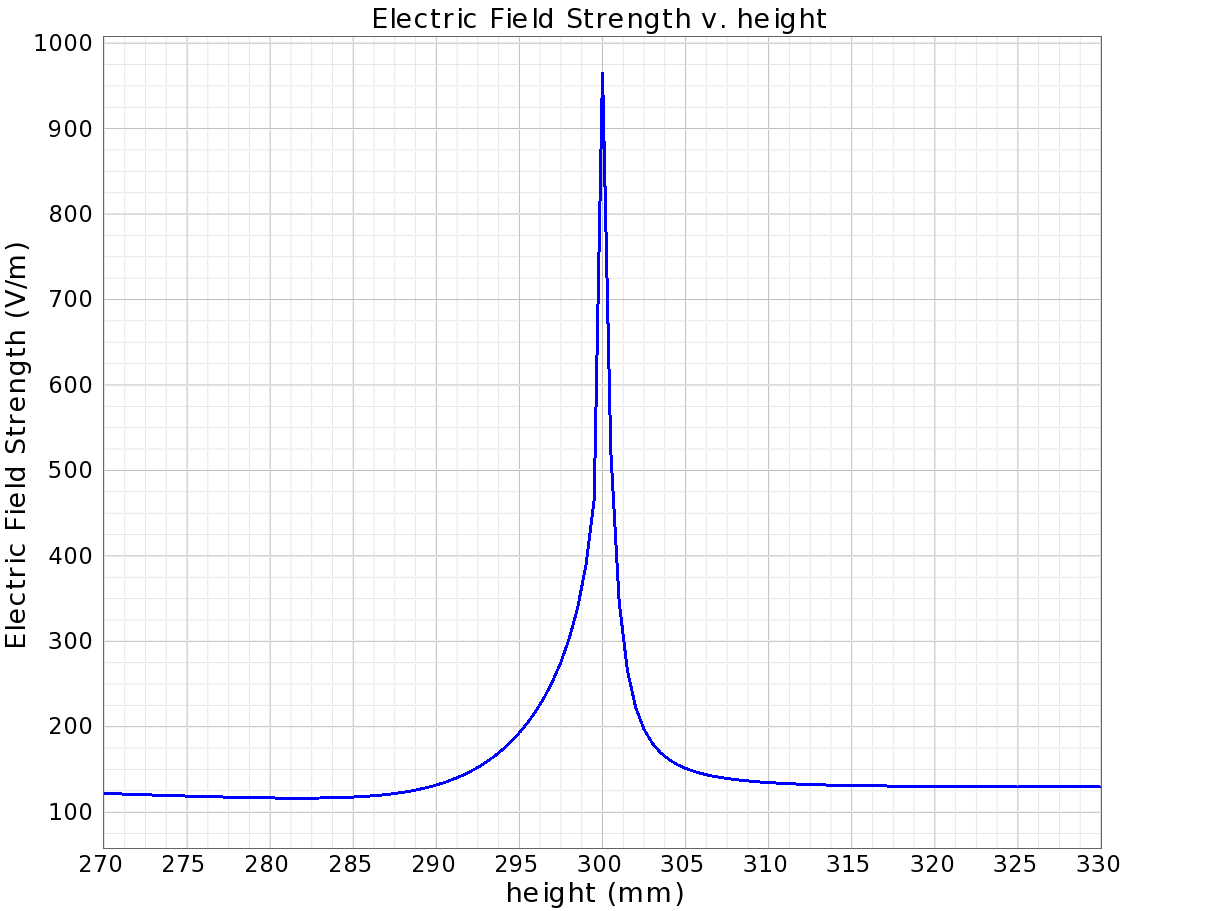}}
\centerline{\includegraphics[width=2.75in]{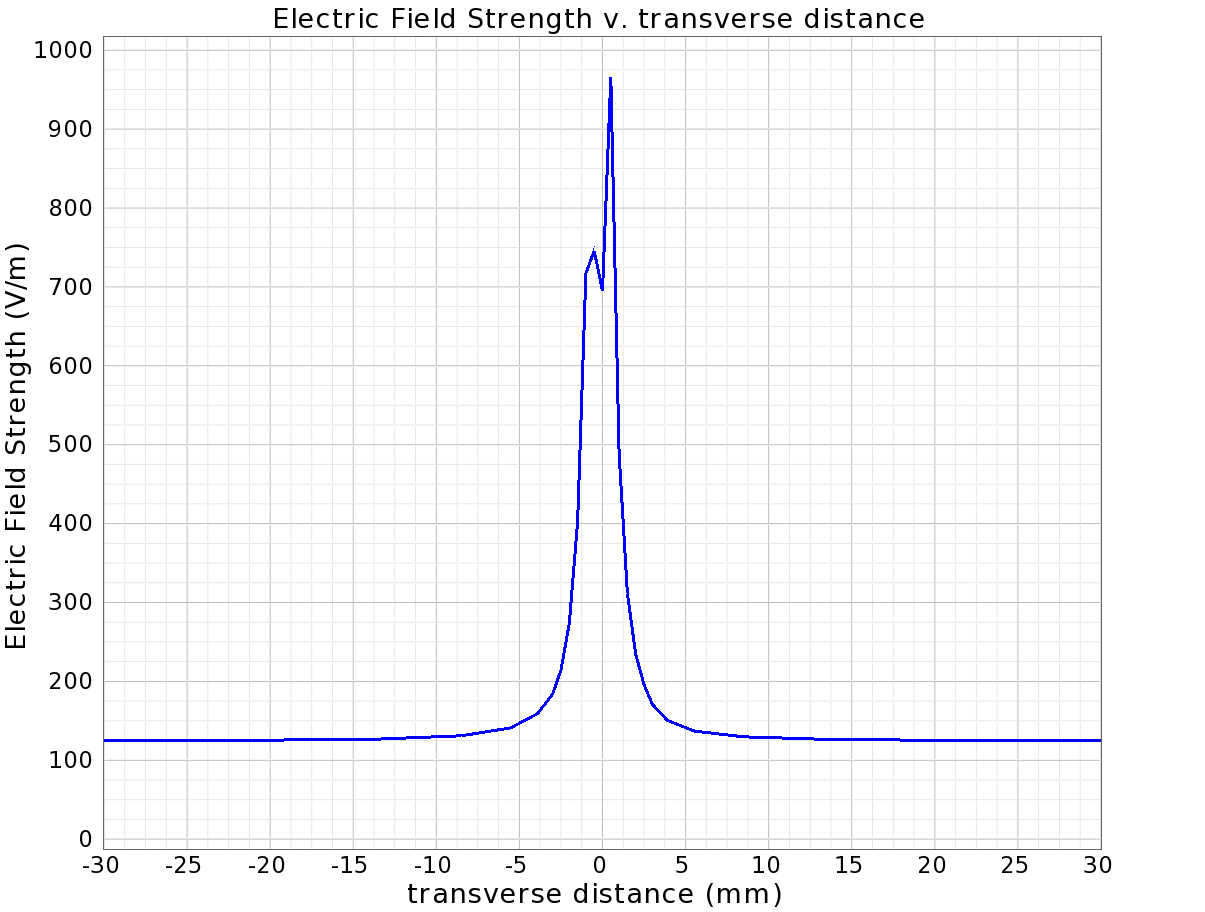}}
\caption{ \footnotesize \it Top: vertical profile of Remcom XFDTD simulation of fair-weather APG fields along a 
simulated plant stem. Bottom: Horizontal profile through the tip of the same stem.
\label{fairAPG} 
}
\end{figure}

We have modeled fair weather APG conditions 
with a plant stem prominence that has parameters similar to plant stems in nature,
with electrical properties determined largely by the water-based electrolytes in their
sap. Whole-stem conductivities for plants fall in a typical range of $\sigma = 10-100$~mS/m, with relative dielectric constants
of order $\epsilon_r =10-50$, again strongly affected by water content~\cite{stemCond1,stemCond2}. We estimate the near-surface 
conductivity, which matters for the surface electric field development, to approach that of saline water:
$\sim 4$~S/m; we use $\sigma = 1$~S/m as the surface conductivity, and $\epsilon_r = 50$, accounting
for surface adsorption of water vapor in a humid environment. The same plate geometry as the 
simulation above is used, except that we use a top plate voltage of 110~V, giving an ambient
field of about 140~V/m, although with 10\% non-uniformity due to the finite plate size.
The plant stem is assumed to be a 30~cm high conical frustum, 0.4~cm in diameter at the bottom and 1~mm in diameter at the top.

Fig.~\ref{fairAPG} shows the results of this simulation, with profiles of the resulting electric fields around
the tip of the plant stem, both in the transverse, and vertical directions.
The small diameter of the dielectric tip strongly magnifies the field, although much less so than a
conducting tip would. The resulting fields approach 1~kV/m in a region several mm around the tip,
with some irregularities due to the mesh structure.
This field appears to be well within the sensor range of the spiders tested previously by LDV methods
when on the cardboard prominence, where fields as low as 400~V/m showed
a clear response well above background noise~\cite{Morley2018}. Thus it appears that at least from a sensory
perspective, these spiders are sensitive to changes at the field levels of Earth's fair weather
APG.

The nanocoulomb charges observed in these experiments would produce very small lift, $\sim2$\% of their weight,
for a spider with a single ballooning silk strand in the fair weather field, however. 
Thus a crucial question arising from these results
is whether spiders modify their behavior and silk extrusion process to adapt to lower fields. 
Certainly one approach is to use longer silk, for example ballooning silks up to 3~m long
have been observed in natural ballooning of larger spiders~\cite{Cho2018}. Assuming a linear
increase in total charge, an order of magnitude more lift may be possible by this method alone
for the spiders considered here.
If in turn another half-dozen or more such silks were extruded, as implied by Fig.~\ref{silkfan},
the {\it Erigone} spiders observed in our experiment would achieve positive buoyancy 
even in the fair weather atmospheric potential gradient, without the aid of any wind.

In conclusion, we have measured the charge state of spider ballooning silk by observing
the acceleration it produced on the spider+silk system in an electric field. 
These accelerations were observed in controlled conditions in which no wind
was present, thus the lift produced, and the resulting launches of the spiders, was
due purely to electrostatic forces on the silk itself, since the spider body charge could not produce
the motion seen. 

These observations, combined with previous measurements
showing that the presence of electrostatic fields elicits ballooning responses, and that
spiders have sensory organs to detect both static and changing electrostatic fields~\cite{Morley2018}, gives 
credence to the proposal that spider ballooning is not a purely aerodynamic process,
but involves an electrostatic component. These results also strongly suggest that spiders
have developed adaptations to directly exploit the Earth's atmospheric potential gradient in
ballooning dispersal behavior; if confirmed they would become the only known organisms of any kind
to make active use of the global electric field.
Further work is still needed to determine 
the interplay between aerodynamic and electrostatic forces, and their relative contributions
under different circumstances.

\begin{acknowledgments}
 The experimental work was funded by a research grant to ELM by the Association for
 the Study of Animal Behavior. ELM would like to thank
 Penelope Fialas for her assistance in data collection and Daniel Robert for providing
 lab space and equipment. 
\end{acknowledgments}


\begin{thebibliography}{99}

\bibitem{MartinLister} Martin Lister, ``Some Observations Concerning the Odd Turn
of Some Shell Snailes, and the Darting of Spiders, made by an Ingenious Cantabrigian and by Way of Letter Communicated
to Mr J. Wray, who Transmitted them to the Publisher for the R.S.'' Phil. Trans. of the Royal Soc. 4 (1669).

\bibitem{Blackwall1827} John Blackwall, Observations and Experiments, made with a view to ascertain the Means by 
which the Spiders that produce Gossamer effect their aërial Excursions, Transactions of the Linnean Society of London,
December 1827,  https://doi.org/10.1111/j.1095-8339.1826.tb00126.x.

\bibitem{Darwin} Charles Darwin,  A Naturalists Voyage: Journal Of Researches Into The Natural History And Geology Of 
The Countries Visited During The Voyage Of H. M. Beagle Round The World Under The Command Of Captain Fitz Roy, R. N.,
Published by John Murray, London (1889)

\bibitem{Duffey56} Duffey, E., Aerial Dispersal in a known spider population, J. Anim. Ecol. 25, 85 (1956).

\bibitem{M3} Bell, JR, Bohan DA, Shaw EM, Weyman GS, Ballooning dispersal using silk:
world fauna, phylogenies, genetics and models., Bull. Entomol. Res. 95, 69-114 (2005).
\bibitem{M4} Weyman, G A, Sunderland, KD, Jepson, PC, review of the evolution and mechanisms
of ballooning by spiders inhabiting arable farmland. Ethol. Ecol. Evol. 14 (4), 307-26 (2002). 
\bibitem{M5} Reynolds A, Beating the odds in the aerial lottery: passive dispersers select 
conditions at takeoff that maximize their expected fitness on landing, Am. Nat. 181, 555-61 (2013). 

\bibitem{Humphrey87}J. A. C. Humphrey, Fluid Mechanic Constraints on Spider Ballooning
Oecologia, Vol. 73, No. 3 (1987), pp. 469-477.


\bibitem{Reynolds2006} Reynolds A.M, Bohan D.A and Bell J.R Ballooning dispersal in arthropod taxa with convergent behaviours: 
dynamic properties of ballooning silk in turbulent flows. Biol. Lett. 2(3), 371-373, (2006).

\bibitem{Reynolds2007} Andy M Reynolds, David A Bohan, James R Bell,
Ballooning dispersal in arthropod taxa: conditions at take-off
Biol Lett. 2007 Jun 22; 3(3): 237-240. Published online 2007 Mar 27. doi: 10.1098/rsbl.2007.0109.


\bibitem{Zhao2017} Zhao L., Panayotova I.N., Chuang A., Sheldon K.S., Bourouiba L., Miller L.A. (2017) Flying Spiders: 
Simulating and Modeling the Dynamics of Ballooning. In: Layton A., Miller L. (eds) 
Women in Mathematical Biology. Association for Women in Mathematics Series, vol 8. Springer, Cham

\bibitem{Cho2018} Moonsung Cho, Peter Neubauer, Christoph Fahrenson, Ingo Rechenberg,
An observational study of ballooning in large spiders: Nanoscale multifibers enable large spiders’ soaring flight,
PLOS Biology 16(6): e2004405. https://doi.org/10.1371/journal.pbio.2004405

%
%
%
%

\bibitem{Schneider2001}
Jutta M. Schneider, J\"org Roos, Yael Lubin, Johannes R. Henschel
DISPERSAL OF STEGODYPHUS DUMICOLA (ARANEAE, ERESIDAE): THEY DO BALLOON AFTER ALL!
Journ. of Arachnology, 29(1):114-116 (2001)

\bibitem{Kronenberger2015} : Kronenberger K, Vollrath F.
2015 Spiders spinning electrically charged
nano-fibres. Biol. Lett. 11: 20140813.
http://dx.doi.org/10.1098/rsbl.2014.0813

\bibitem{Perolo2017}
A. Perolo, A. Castiglioni, and D. Ferri,
Electrification of polymers during capillary
extrusion, AIP Conference Proceedings 1914, 040007 (2017); https://doi.org/10.1063/1.5016717


\bibitem{Gorham} Gorham, Peter W., Ballooning Spiders: The Case for Electrostatic Flight, arXiv:1309.4731 [physics.bio-ph], (2013).


\bibitem{Morley2018} Morley EL \& Robert D., Electric Fields Elicit Ballooning in Spiders,  Curr. Biol 28, 2324-30 (2018). 





\bibitem{M7} Wilson, CTR, Atmospheric Electricity, Nature 68(1753), 102-104 (1903) 

\bibitem{Wilson1920} Wilson, C.T.R., 1921. Investigations on lightning discharges and the electric
 field of thunderstorms. Phil. Trans. A. 221, 73-115.

 
\bibitem{M6} Michael J. Rycroft,Keri A. Nicoll, Karen L. Aplin, R. Giles Harrison, 
Recent advances in global electric circuit coupling between the space
environment and the troposphere J. Atmos. Sol.-Terr. Phys. 90-1, 198-211 (2012). 


\bibitem{Barth2004} Barth, F., Spider Mechanoreceptors, Current Opinion in Neurobiology, 14(4), 415-422,(2004).


\bibitem{houseflyTribo} Daniel F. McGonigle, Chris W. Jackson, John L. Davidson,
Triboelectrification of houseflies
(Musca domestica L.) walking on synthetic
dielectric surfaces,
Journal of Electrostatics 54 (2002) 167-177

\bibitem{beeCharge1} 
M.E. Colin, D. Richard, S. Chauzy, Measurement of electric charges carried by bees: evidence of
biological variations, J. Bioelectric. 10 (1) (1991) 17-32;

\bibitem{beeCharge2} 
Uwe Greggers, Gesche Koch , Viola Schmidt , Aron D\"urr , Amalia Floriou-Servou , David Piepenbrock , Martin C. G\"opfert  and Randolf Menzel,
Reception and learning of electric fields in bees,
Proc. of the Royal Society B 280, 20130528 (2103), \url{http://doi.org/10.1098/rspb.2013.0528}

\bibitem{beeCharge3}Colin ME, Richard D, and Chauzy S. (1991),
Measurement of electric charges carried by bees: evidence of biological variations. J. Bioelectr. 10, 17–32. 


\bibitem{beeArea}
Yannick Poquet,Laurent Bodin,Marc Tchamitchian,Marion Fusellier,Barbara Giroud,
Florent Lafay,Audrey Bulet\'e,Sylvie Tchamitchian,Marianne Cousin,Michel P\'elissier,Jean-Luc Brunet,Luc P. Belzunces
A Pragmatic Approach to Assess the Exposure of the Honey Bee (Apis mellifera) When Subjected to Pesticide Spray
PLOS ONE 9(11), (2014), e113728. https://doi.org/10.1371/journal.pone.0113728

\bibitem{Remcom}Remcom Inc.,  XFdtd 3D Electromagnetic Simulation Software, \url{https://www.remcom.com/xfdtd-3d-em-simulation-software}.

\bibitem{CompPhys}Rubin H. Landau, Manuel J P\'aez, Cristian C. Bordeianu,  Computational Physics: Problem Solving with Python,
3rd Edition, (Wiley VCH),
ISBN: 978-3-527-41315-7 September 2015 644 Pages

\bibitem{Penell2018} Anja Penell, Florian Raub, Hubert H\"ofer,
"Estimating biomass from body size of European spiders based on regression models," 
The Journal of Arachnology, 46(3), 413-419, (1 November 2018).









\bibitem{silkCond1} Steven E, Park JG, Paravastu A, Lopes EB, Brooks JS, Englander O, Siegrist T, Kaner P, Alamo RG. 
Physical characterization of functionalized spider silk: electronic and sensing properties. 
Sci Technol Adv Mater. 2011 Aug 23;12(5):055002. doi: 10.1088/1468-6996/12/5/055002.

%

\bibitem{Suter92} R. B. Suter, Ballooning: Data from spiders in freefall indicate the importance of posture,
The Journal of Arachnology 20:107-113 (1992).



\bibitem{Erickson09} H P Erickson, Size and shape of Protein Molecules at the Nanometer Level Determined by
Sedimentation, Gel Filtration, and Electron Microscopy, Biol. Proced. Online (2009), 11, 32-51.


\bibitem{SpaceCharge98} R. Gerhard-Multhaupt, W. K\"unstler, G. Eberle, W. Eisenmenger and G. Yang, High
space-charge densities in the bulk of fluoropolymer electrets detected with piezoelectrically generated 
pressure steps. In: J. C. Fothergill and L. A. Dissado (eds.), Space
Charge in Solid Dielectrics, pp. 123–132, Dielectrics Society, Leicester, England
(1998).

\bibitem{highSC2018} Francesco Bonacci, Alessandro Di Michele, Silvia Caponi, Francesco Cottone and Maurizio Mattarelli, 
High charge density silica micro-electrets fabricated by electron beam, 
Smart Mater. Struct. 27 075052, (2018).


\bibitem{spaceCharge} B\"oer, Karl. (2010). Introduction to Space Charge Effects in Semiconductors,
Springer Series in Solid-State Sciences, doi:10.1007/978-3-642-02236-4. 

\bibitem{flowElect}
F. Flores , D. Graebling , A. Allal and C. Guerret-Pi\'ecourt,
Modelization of flow electrification in a polymer melt,
J. Phys. D 40, 2911 (2007).

\bibitem{flowElect2}
G\'erard G. Touchard, Tadeusz W. Patzek, and Clayton J. Radke,
A Physicochemical Explanation for Flow
Electrification in Low-Conductivity
Liquids in Contact with a Corroding Wall,
IEEE TRANSACTIONS ON INDUSTRY APPLICATIONS, VOL. 32, NO. 5, 1051, (1996).

%
%
%
%


\bibitem{stemCond1} Nadler, A. Raveh, E. Yermiyahu, U. Lado, M., Nasser A., Barak, M., and Green, S.,
Detecting Water Stress in Trees Using Stem Electrical Conductivity Measurements,
Soil Sci. Soc. Am. J. 72:1014-1024, (2008),
doi:10.2136/sssaj2007.0308

\bibitem{stemCond2} Eunyong JEON, Sangwoong BAEK, Seungyul CHOI, Kyoung Sub PARK, Junghoon LEE, 
Real-Time Monitoring of Electroconductivity in Plants with Microscale Needle Probes, 
Environmental Control in Biology Volume 56 (2018) Issue 4, 131-135. \url{https://doi.org/10.2525/ecb.56.131}




\end{thebibliography}
\end{document}